\documentclass[a4paper,11pt]{article}
\pdfoutput=1 % if your are submitting a pdflatex (i.e. if you have
\usepackage{jinstpub} % for details on the use of the package, please
\usepackage{lineno}
\title{\boldmath Multi-detector approach to enhance the sensitivity of neutrino telescopes to low-energy astrophysical sources}

\author[a]{Gwenha\"el de Wasseige\note{Corresponding author.}}
\affiliation[a]{Laboratoire APC, Universit\'e de Paris\\10, Rue Alice Domon et L\'eonie Duquet, 75013 Paris, France}

\emailAdd{gdewasseige@km3net.de}

\abstract{While large neutrino telescopes have so far mainly focused on the detection of TeV-PeV astrophysical neutrinos, several efforts are ongoing to extend the sensitivity down to the GeV level for transient sources. Only a handful of neutrino searches have been carried out at the moment leaving the signature of astrophysical transients poorly known in this energy range.
In this contribution, we discuss the motivations for high-energy neutrino telescopes to explore the GeV energy range and summarize the current limitations of detectors, such as IceCube and KM3NeT. We then present and compare different approaches for multi-detector analyses that may enhance the sensitivity to a transient GeV neutrino flux.}

\proceeding{Very Large Volume Neutrino Telescope 2021\\
18-21 May 2021\\
Valencia, Spain}

\begin{document}
\maketitle
\flushbottom

\section{Motivation}
\label{sec:intro}
\begin{figure}[htbp]
\centering % \begin{center}/\end{center} takes some additional vertical space
\includegraphics[width=.4\textwidth]{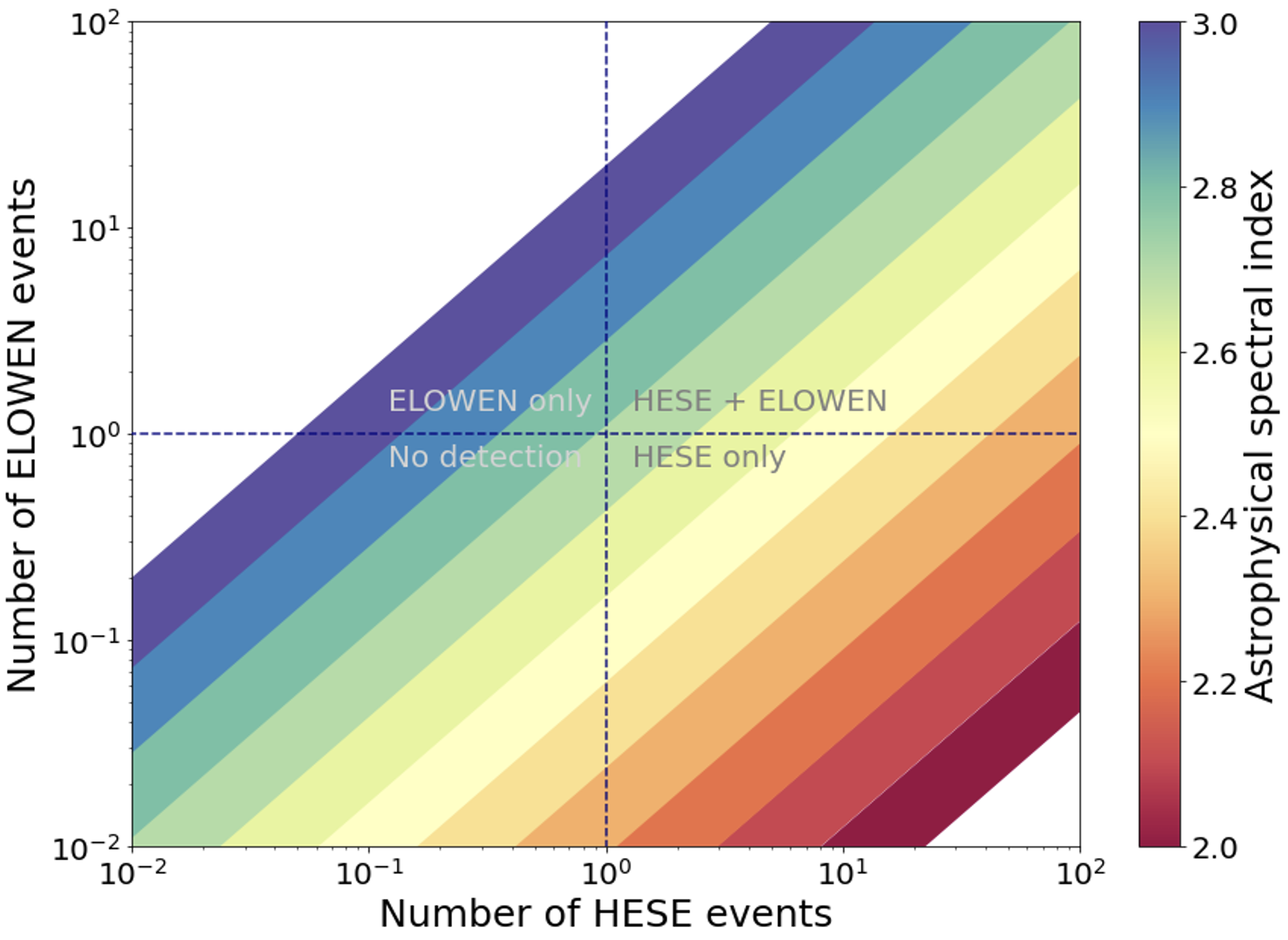}
\qquad
\includegraphics[width=.4\textwidth]{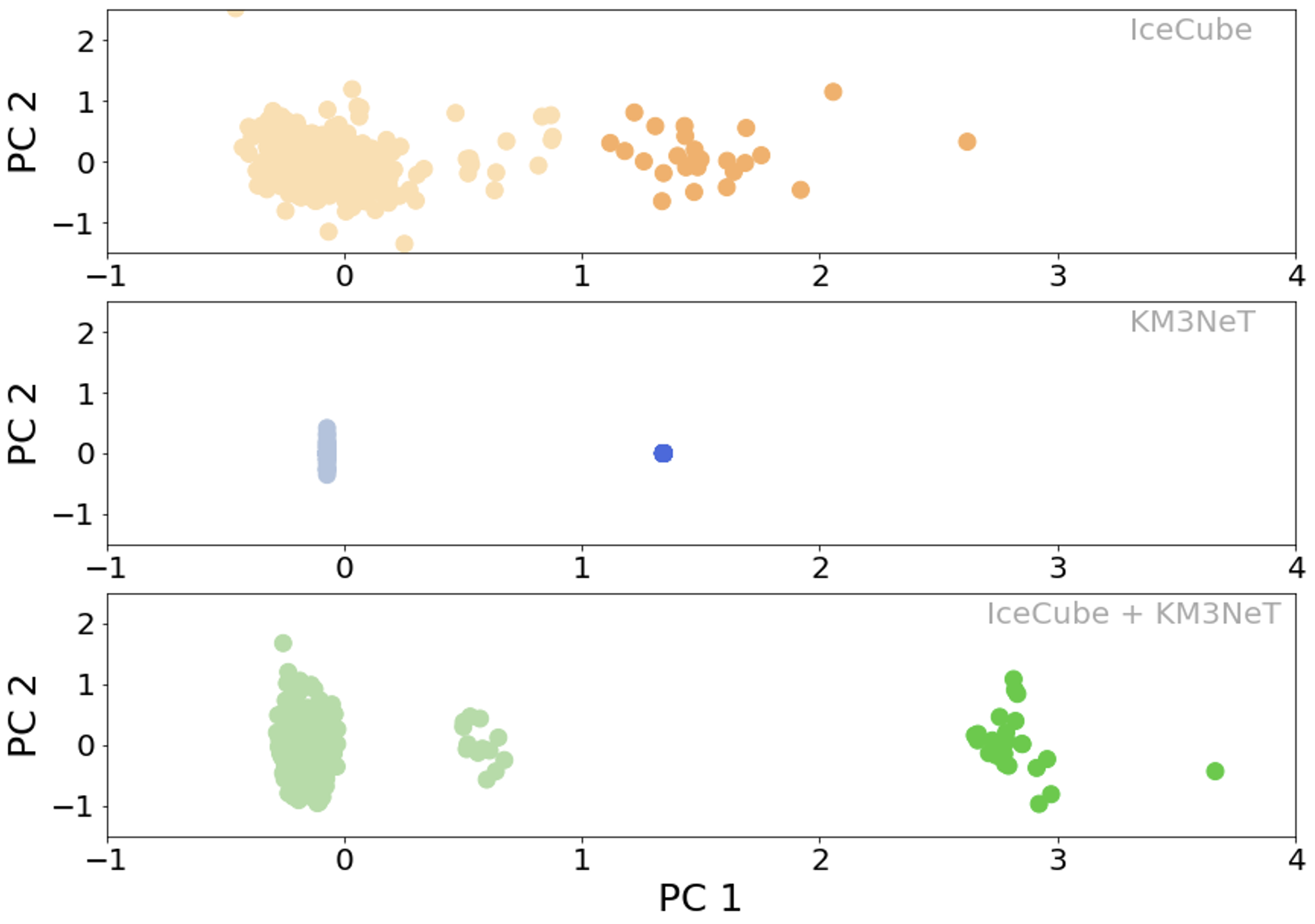}
%% "\includegraphics" from the "graphicx" permits to crop (trim+clip)
%% and rotate (angle) and image (and much more)
\caption{\label{fig:lowvshigh} Left panel : Number of detected low-energy events (between 500~MeV and 5~GeV) in IceCube as function of the number of HESE events for different values of the astrophysical spectral index when assuming a non-broken power law between the two considered energy ranges. Right: Results obtained when applying PCA to the pseudo-experiments in IceCube (top), KM3NeT (middle), and the combination (bottom) when injecting signal events in 5\% of the considered population. In each panel, the darker color represents the pseudo-experiments in which signal events were added}
\end{figure}

Very large neutrino telescopes, such as IceCube and ANTARES, have mainly searched so far for astrophysical neutrinos at the TeV energy and above, leaving the sub-TeV regime poorly explored with only a few searches carried out (e.g.,~\cite{sf,gw,michael-icrc,greco-untrigger}).
Recent efforts allowed IceCube to be sensitive to a transient neutrino emission  down to 500~MeV~\cite{sf}. In the Mediterranean Sea, the ORCA site of the KM3NeT detector, currently being deployed off-shore Toulon in France, is optimized to detect GeV neutrinos and thus appears as an ideal tool for sub-TeV neutrino astronomy~\cite{nmo-astro,online}. Both instruments will be able to produce in the coming years useful constraints/observations of the sub-TeV neutrino flux coming from promising source populations as binary compact mergers~\cite{gw}, novae~\cite{michael-icrc}, solar flares~\cite{sf}, and carry out untriggered transient searches~\cite{greco-untrigger}.
Both instruments suffer from a background due to the detector itself and its environment representing for now the limiting factor for an enhanced sensitivity. In this contribution, we present preliminary investigations on methods to combine IceCube and KM3NeT data in order to boost the sensitivity in the GeV energy range. While the direction reconstruction of low-energy neutrino interactions is challenging and poorer than what is feasible at higher energies, IceCube and KM3NeT large volumes would guarantee a large number of interactions to be recorded. For illustration purpose, Fig.~\ref{fig:lowvshigh} (left panel) shows the expected neutrino events expected using the ELOWEN event selection in IceCube~\cite{sf}, sensitive between 500~MeV and 5~GeV, as function of the number of HESE events~\cite{hese} assuming a non-broken power-law from the GeV to the PeV range. The most recent estimate of the high-energy astrophysical neutrino flux detected by IceCube being around 2.8~\cite{hese}, more than 1 GeV neutrino would be detected for each HESE event if it were coming from a transient source. Low-energy searches therefore appear as a complementary way to probe the neutrino emission in transient sources.
\section{Description of the adopted approach and results of preliminary investigations}
In order to mitigate the impact of the background, we propose to combine data recorded at different sites. Background events produced by atmospheric muons and neutrinos as well as environmental or detector noise would be uncorrelated while an astrophysical signal may appear causally connected in the two sites. Let us consider as an example the follow-up searches that would be carried out during the fourth run (O4) of observation of the LIGO, Virgo, and KAGRA interferometers. Motivations to search for a GeV neutrino flux from compact binary mergers have been presented in e.g.,~\cite{imre, gw}. IceCube has set upper limits of the order of $10^{4}$ $\nu$ cm$^{-2}$~\cite{gw} from such sources detected during the first three runs of observation of LIGO and Virgo. The goal in this contribution is to check whether one could identify a sub-population, emitting an arbitrary low GeV neutrino flux, among the mergers detected during O4 by combining IceCube and KM3NeT data. 

We assume 500 compact binary mergers will be detected during O4 and follow-up searches will be carried out in the 1000~s time window around the merger time. A naive approach is to add the total number of events detected in IceCube and KM3NeT during those 1000~s. A Kolmogorov-Smirnov test can then be performed to compare the distribution obtained after the 500 follow-ups and the background-only distribution. This approach requires at least 20\% of the compact binary mergers to be emitting a flux at Earth larger than $6\times 10^{2}$ $\nu$ cm$^{-2}$ to lead to a significant observation.

In view of improving the sensitivity, we propose to add the time information and use a Primary Component Analysis (PCA) to find a potential source population. We divide the 1000~s in 1000 bins of 1~s and sum the number of events detected in IceCube and KM3NeT in each of these bins. We run 500 pseudo-experiments for each detector, taking into account the expected background rate, 20~mHz and $3.4 \times 10^{-2}$ ~mHz for IceCube~\cite{sf} and KM3NeT~\cite{nmo-astro}, respectively.
Using the elbow method~\cite{datamining}, we verify that PCA does not find any subgroup in the pseudo follow-ups when only background events are generated.
%no clear elbow is found in the distribution of the inertia as function of the number of clusters indicating that the most likely number of clusters in the distribution is one.
We then inject signal events in 5\% of the pseudo follow-ups. We start by adding one event in each detector at a bin N and N+1 and verify if PCA can identify the pseudo-experiments in which the signal was added. Figure~\ref{fig:lowvshigh} (right panel) shows the obtained results for IceCube (top), KM3NeT (middle), and the combination (bottom) where the darker color in each panel represents the pseudo-experiments with injected signal.  The elbow method indicates that the most likely number of clusters in the distribution is two: the background-only pseudo-experiments and those where the signal events were added. This result confirms the efficiency of the method. 
We have reproduced the experiment assuming different hypotheses for the injected signal, such as two sub-populations with a precursor and a prompt signal injected. The proposed method was in every case able to identify the correct number of sub-populations. 

\begin{figure}[htbp]
\centering % \begin{center}/\end{center} takes some additional vertical space
\includegraphics[width=.4\textwidth]{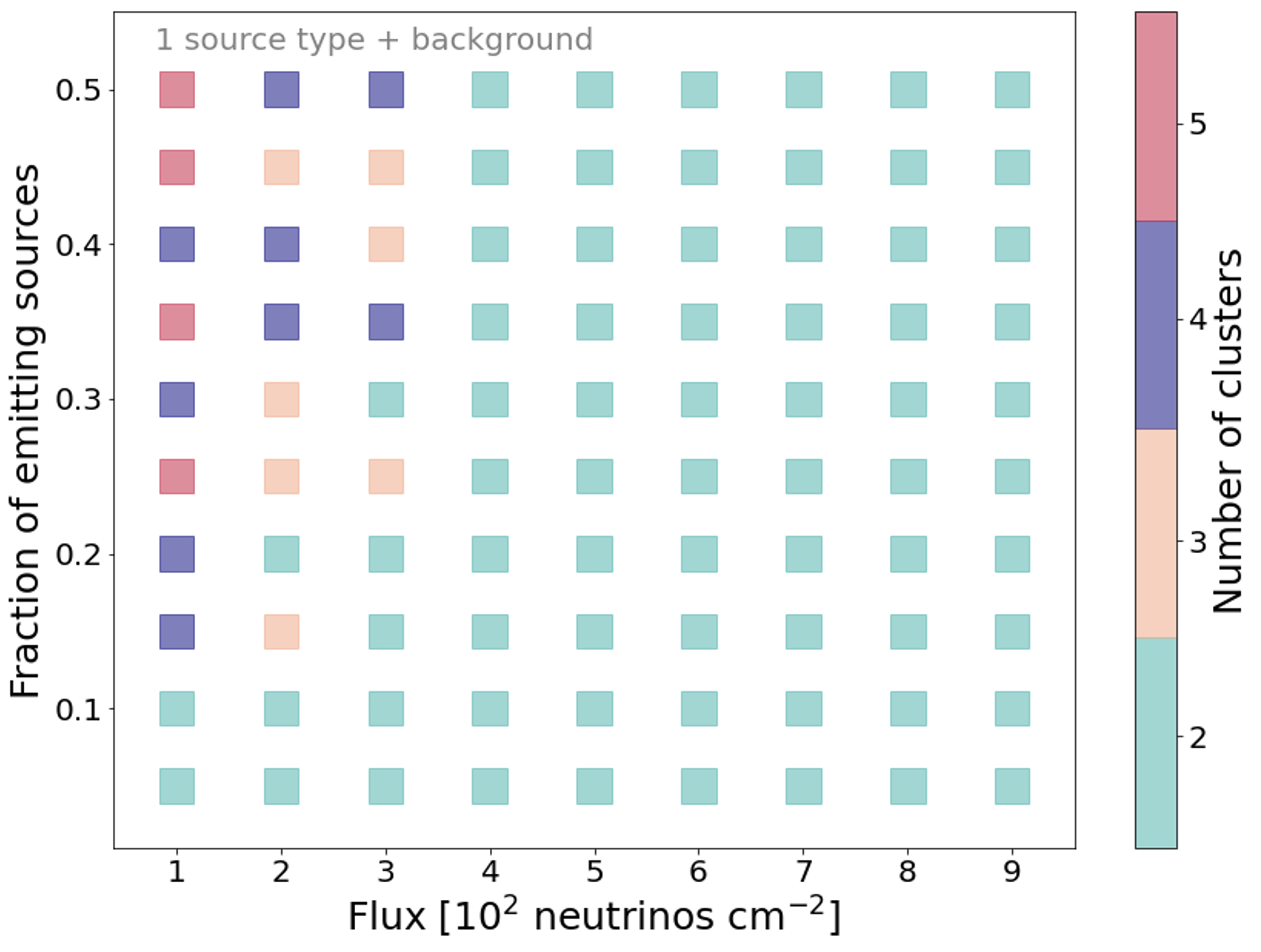}
\qquad
\includegraphics[width=.4\textwidth]{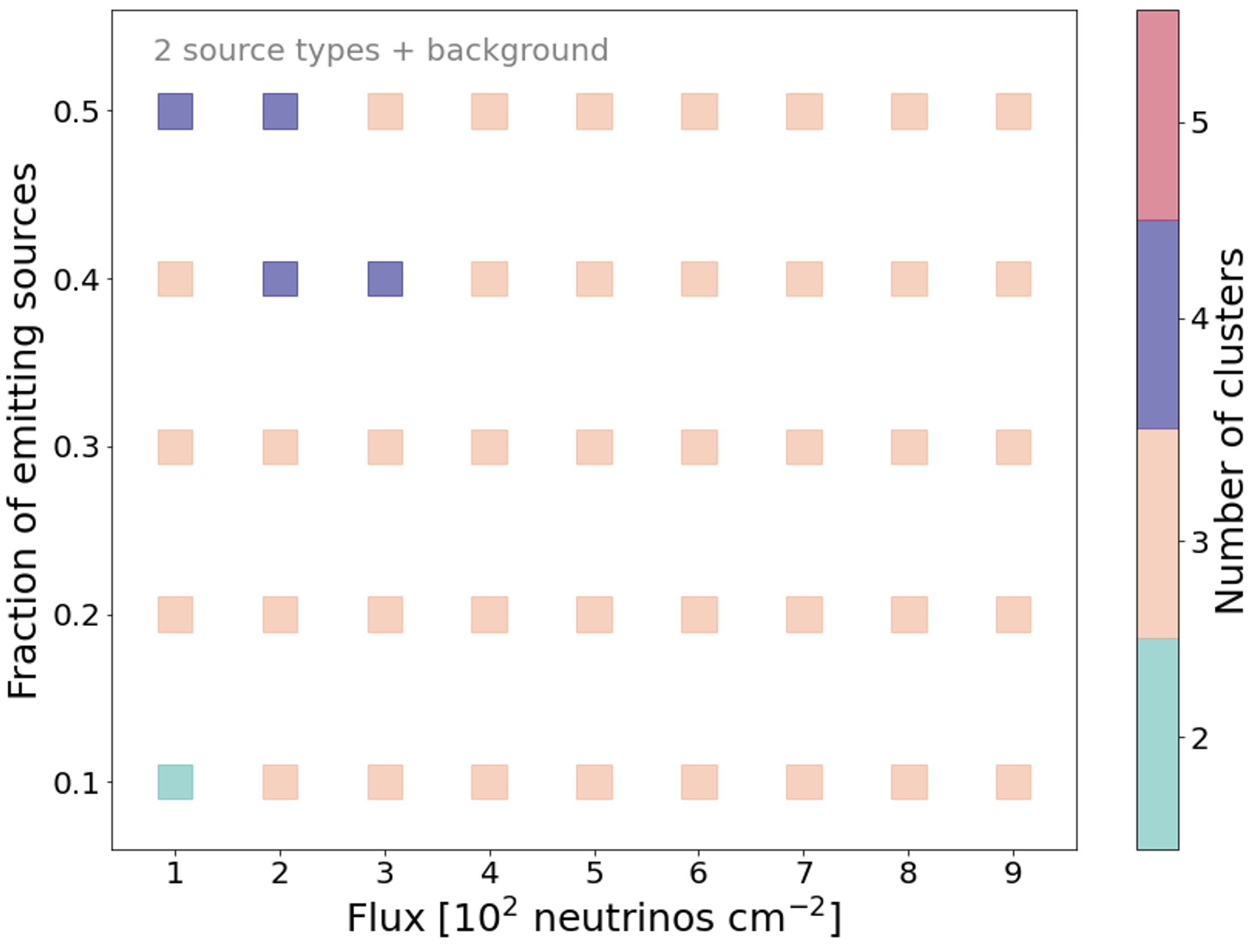}
%% "\includegraphics" from the "graphicx" permits to crop (trim+clip)
%% and rotate (angle) and image (and much more)
\caption{\label{fig:pca-general} 	Results obtained when applying PCA to pseudo-experiments in the combination of IceCube and KM3NeT with one sub-population (left) and two different sub-populations (right) of GeV-neutrino emitters. The x-axis represents the injected signal in terms of astrophysical flux at Earth and the y-axis shows the fraction of the merger population in which a signal has been injected. The color code represents the number of sub-populations found by the method. }
\end{figure}

In order to generalize the methods, we run pseudo-experiments with injected signal events following a Poisson distribution with a mean between 0.5 and 5  events in a fraction of the compact binary mergers between 5\% and 50\% of the 500 assumed sources. For each case we evaluate the number of clusters found in the distribution. The results are show in Fig.~\ref{fig:pca-general} for one sub-population with injected signal (left) and two different sub-populations (precursor and prompt signal events, right).
For small sub-populations, i.e. < 15\% of the mergers, the approach successfully finds the exact number of sub-populations when we inject a flux as small as $1\times 10^{2}$ $\nu$ cm$^{-2}$, while for larger populations, a flux of $4\times 10^{2}$ $\nu$ cm$^{-2}$ is needed to find the exact number of assumed sub-populations. The proposed method therefore leads to an improvement compared to the naive approach previously described.

%\begin{figure}[htbp]
%\centering % \begin{center}/\end{center} takes some additional vertical space
%\includegraphics[width=.4\textwidth]{pca}
%% "\includegraphics" from the "graphicx" permits to crop (trim+clip)
%% and rotate (angle) and image (and much more)
%\caption{\label{fig:pca} Results obtained when applying PCA to the pseudo-experiments in IceCube (top), KM3NeT (middle), and the combination (bottom) when injecting signal events in 5\% of the considered population. In each panel, the darker color represents the pseudo-experiments in which signal events were added.}
%\end{figure}

\section{Summary and outlook}
Several very large neutrino telescopes are now sensitive in the sub-TeV range. While each instrument can perform searches alone, the obtained sensitivity suffers from the presence of an almost irreducible background. The combination of observations from different sites would reduce the impact of the background and enhance the significance of an astrophysical signal. We propose a method using a Principal Component Analysis and apply it to follow-ups searches that would be done during the fourth run of observation of LIGO, Virgo, and KAGRA interferometers as a test scenario.
Our first investigation seems promising in finding sub-populations of GeV-neutrino emitters within the full population of the binary compact mergers. We plan on continuing investigating different techniques allowing us to enhance the sensitivity in this poorly explored energy range.

\acknowledgments
G. de Wasseige acknowledges support from the European Union's Horizon 2020 research and innovation programme under the Marie Sklodowska-Curie grant agreement No 844138.


\begin{thebibliography}{99}
%\bibitem{loi}
%{\bf KM3NeT} Collaboration, Letter of intent for KM3NeT 2.0, J. Phys. G: Nucl. Part. Phys. {\bf 43} (2016) 084001.
%\bibitem{sn_paper}
%{\bf KM3NeT} Collaboration, Eur. Phys. J. C {\bf 81} (2021) 445.

\bibitem{sf}
{\bf IceCube} Collaboration, R. Abbasi \textit{et al.}, %The Design and Performance of IceCube DeepCore, 
 \href{https://journals.aps.org/prd/abstract/10.1103/PhysRevD.103.102001}{Phys. Rev. D {\bf103}, 102001 (2021)}.

\bibitem{gw}
{\bf IceCube} Collaboration, R. Abbasi \textit{et al.}, %The Design and Performance of IceCube DeepCore, 
 \href{https://arxiv.org/abs/2105.13160}{arXiv:2105.13160}.
  
\bibitem{michael-icrc}
M. Larson for the {\bf IceCube} Collaboration, \href{https://pos.sissa.it/395/1131/pdf}{PoS(ICRC2021)1131}.

\bibitem{greco-untrigger}
{\bf IceCube} Collaboration, R. Abbasi \textit{et al.},  \href{https://arxiv.org/abs/2011.05096}{arXiv:2011.05096}.

\bibitem{nmo-astro}
G. de Wasseige for the {\bf KM3NeT} Collaboration, \href{https://pos.sissa.it/358/934/pdf}{PoS(ICRC2019)934}.

\bibitem{online} 
F. Huang for the {\bf KM3NeT} Collaboration, Real-time Multi-Messenger Analysis Framework of KM3NeT, these proceedings, \href{https://arxiv.org/abs/2107.13908}{arXiv:2107.13908}.

\bibitem{hese}
{\bf IceCube} Collaboration, R. Abbasi \textit{et al.}, 
 \href{https://journals.aps.org/prd/abstract/10.1103/PhysRevD.104.022002}{Phys. Rev. D 104, 022002 (2021)}.
\bibitem{imre}
K. Asano, K. Murase, \href{https://doi.org/10.1155/2015/568516}{Adv. Astron., 568516 (2015)}.

\bibitem{datamining}
M. Zaki and W. Meira, \textit{Data Mining and Analysis: Fundamental Concepts and Algorithms}, Cambridge University Press. March 2020, ISBN: 978-1108473989.
\end{thebibliography}
\end{document}